\documentclass[prl,aps,preprint]{revtex4}%
\usepackage[dvips]{graphicx}
\usepackage{dcolumn}%
\usepackage{amsmath}%
\usepackage{amsfonts}%
\usepackage{amssymb}
 \usepackage[usenames]{color}
\providecommand{\U}[1]{\protect\rule{.1in}{.1in}}
\newcommand{\be}{\begin{equation}}
\newcommand{\ee}{\end{equation}}
\newcommand{\bea}{\begin{eqnarray}}
\newcommand{\eea}{\end{eqnarray}}
\newcommand{\bt} {\begin{tabular}}
\newcommand{\et} {\end{tabular}}
\newcommand{\nn}{ \nonumber}
\newcommand{\ds}{\displaystyle}
\newcommand{\ba} {\begin{array}}
\newcommand{\ea} {\end{array}}
\begin{document}

\title{Energy, work, entropy and heat balance in Marcus molecular junctions}

\author{  Natalya A. Zimbovskaya$^1$ and Abraham Nitzan$^{2,3}$}

\affiliation
{Department of Physics and Electronics, University of Puerto Rico-Humacao, CUH Station, Humacao, PR 00791, USA}  
\affiliation{$^2$Department of Chemistry, University of Pennsylvania, Philadelphia, PA19104, USA}
\affiliation{$^3$School of Chemistry, Tel Aviv University, Tel Aviv, Israel}

\begin{abstract}
We present a consistent theory of energy balance and conversion in a single-molecule junction with strong interactions between electrons on the molecular linker (dot) and phonons in the nuclear environment where the Marcus-type electron hopping processes predominate in the electron transport. It is shown that the environmental  reorganization and relaxation that accompany electron hopping energy exchange between the electrodes and the nuclear (molecular and solvent) environment may bring a moderate local cooling of the latter in biased systems. The effect of a periodically driven dot level on the heat transport and power generated in the system is analyzed and energy conservation is demonstrated both within and beyond the quasistatic regime. Finally, a simple model of atomic scale engine based on a Marcus single-molecule junction with a driven electron level is suggested and discussed.
\end{abstract}

\date{\today}
\maketitle

\subsection{I. Introduction} 

In the last two decades molecular electronics became a well established and fast developing field \cite{1,2,3,4,5}. Presently, it provides a general platform which can be used to consider diverse nanoscale electronic and energy conversion devices. The basic building block of such devices is a single-molecule junction (SMJ). This system consists of a couple of metallic/semiconducting electrodes linked with a molecular bridge. Electron transport through SMJs is controlled by electric forces, thermal gradients and electron-phonon interactions. In addition, in SMJs operating inside dielectric solvents, transport properties may be strongly affected by the solvent response to the molecule charge states \cite{6,7,8}. Overall, electrons on the molecule may interact with a collection of thermalized phonon modes associated with the solvent nuclear environment as well as with individual modes associated with molecular vibrations. Such interactions lead to the energy exchange between traveling electrons and the environment thus giving rise to inelastic effects in the electron transport. In the weak electron-phonon coupling limit inelastic contributions may be treated as perturbations of basically elastic transport \cite{9,10,11,12,13,14,15,16}. Stronger coupling of electrons to phonon modes can result in several interesting phenomena including negative differential conductance, rectification and Franck-Condon blockade \cite{17,18,19,20,21,22}.

In the present work we consider a limit of very strong electron-phonon interactions when electron transport may be described as a sequence of hops between the electrodes and the bridge sites and/or among the bridge sites subjected to local thermalization. This dynamics is usually described by successive Marcus-type electron transfer processes \cite{3,23,24,25,26,27}. Indeed, Marcus theory has been repeatedly and successfully used to study charge transport through redox molecules \cite{28,29,30,31,32,33,34,35}. Nuclear motions and reorganization are at the core of this transport mechanism. The theory may be modified to include effects of temperature gradients across a SMJ \cite{36,37} as well as a finite relaxation time of the solvent environment. Further generalization of Marcus theory accounting for finite lifetime broadening of the molecule electron levels was recently suggested \cite{38,39}.

Besides charge transport, electron-phonon interactions may strongly affect heat generation and transport through SMJs \cite{9,10,20}. There is an increasing interest in studies of vibrational heat transport in atomic scale systems and in their interfaces with bulk substrates \cite{40,41,42,43,44,45}. Electron transfer induced heat transport was also suggested and analyzed \cite{36,37}. Effects of structure-transport correlations on heat transport characteristics of such systems are being studied \cite{46,47,48} as well as effects originating from specific features of coupling between the molecular linker and electrodes \cite{49,50,51,52} and the heat currents rectification \cite{53,54,55}. Correlations between structure and heat transfer in SMJs and similar systems may be accompanied by heating/cooling of the molecular bridge environment \cite{9,10,56,57,58,59,60,61}.

Nevertheless, the analysis of heat transfer in SMJs is far from being completed, especially in molecular junctions dominated by Marcus-type electron transfer processes. In the present work, we theoretically analyze energy balance and conversion in such systems. For the molecular bridge we use the standard single single level model that describes two molecular electronic states in which the level is either occupied or unoccupied. The electrodes are treated as free electron reservoirs with respective chemical potentials and temperatures. We assume that the level may be slowly driven by an external agent (such as a gate voltage) which moves it over a certain energy range. Also, we assume that the temperature of solvent environment  of the bridge may differ from the electrodes temperatures. Despite its simplicity, the adopted model captures essential physics of energy conversion in such junctions. 

The paper is organized as follows. In Sec.II we review the application of Marcus rates to the evaluation of steady state currents resulting from voltage and temperature bias across the junction. We study the relationship between heat currents flowing into the electrodes and into the solvent environment of the molecular bridge and demonstrate overall energy conservation. In Sec.III we analyze the energy balance in a system where a bridge electronic level is driven by an external force. We discuss the irreversible work thus done on the system and the  corresponding dissipated power and the entropy change. In Sec.IV, we describe a simple model for a Marcus junctione engine and estimate its efficiency. Our conclusions are given in Sec.V.

\subsection{II. Steady state currents}
\subsection{A. Electron transfer rates and electronic currents}

 We consider a molecular junction were electrons move between electrodes through a molecular bridge (or dot) that can be occupied ( state $a$) or unoccupied (state $b$). Adopting the Marcus formalism we assume that each state corresponds to a free energy surface which is assumed to be parabolic in the collective solvent coordinate $x$. Here and below we take "solvent" to include also the intramolecular nuclear motion which contributes to the electronic charge accumulation. We use the simplest shifted surfaces model: in terms of $x$, the energy surfaces associated with the two electronic states are assumed to take the forms of identical harmonic surfaces that are shifted relative to each other:
\be
E_a (x) = \frac{1}{2} k x ^2 ,  \label{1}
\ee 
\be
E_b (x) = \frac{1}{2} k (x - \lambda)^2 + \epsilon_d.  \label{2}
\ee
Here, $\lambda$ represents a shift in the equilibrium value of the reaction coordinate and $\epsilon_d$ is the difference between the equilibrium energies of the two electronic states. Diverse reaction geometries may be taken into account by varying $\epsilon_d$ and the force constants \cite{62,63,64}. More sophisticated models \cite{65} make the mathematics more involved but not expected to change the essential physics. The reorganization energy $E_r$ associated with the electron transfer process
\be
E_r = \frac{1}{2} k \lambda^2.   \label{3} 
\ee
reflects the strength of interactions between electrons on the bridge and the solvent environment. For $E_r=0$ electron transport becomes elastic. 

 The overall kinetic process is determined by the transfer rates 
$ k_{a\to b}^{L,R} $ and $ k_{b \to a}^{L,R} $ that correspond to transitions at the left (L) and right (R) electrodes between the occupied ($a$) and unoccupied ($b$) molecular states (namely, $a\to b$ corresponds to electron transfer from molecule to metal and $b\to a$ denotes the opposite process). Because of the timescale separation between electron and nuclear motions, these transfer processes have to satisfy electronic energy conservation: 
\be
g(x,\epsilon) = E_b(x) - E_a(x) + \epsilon = 0.  \label{4}
\ee
where $\epsilon$ is the energy of electron in the metal. This leads to the Marcus electron transfer rates given by \cite{27}:
\be
k_{a \to b}^K = \sqrt{\frac{\beta_s}{4\pi E_r}} \int_{-\infty}^\infty d \epsilon\Gamma_K(\epsilon) [1 - f_K (\beta_K, \epsilon)] \exp \left[-\frac{\beta_s}{4 E_r} (\epsilon + E_r - \epsilon_{d})^2 \right]   \label{5},
\ee
\be
k_{b \to a}^K = \sqrt{\frac{\beta_s}{4\pi E_r}} 
\int_{-\infty}^\infty d \epsilon \Gamma_K(\epsilon)f_K (\beta_K, \epsilon)
\exp\left[- \frac{\beta_s}{4E_r} (\epsilon_{d} + E_r - \epsilon)^2 \right],   \label{6}
\ee
where $K=\{L,R\}$ stands for the left and right electrode. In these expressions, $ \Gamma_{L,R} $ are the bare electron transfer rates between the single molecule level and the electronic continuum in the metal, $ \beta_{L,R} = (kT_{L,R})^{-1} $ and $ \beta_s = (kT_s)^{-1} $ indicate the temperatures of the electrodes and the molecule environment, $k$ is the Boltsmann constant and $ f_{L,R} $ are Fermi distribution functions for the electrodes with chemical potentials $ \mu_{L,R}. $ Expressions Eqs (\ref{5}), (\ref{6}) assume that electron transfer takes place from an equilibrium solvent and metal configurations, namely that thermal relaxation in the metal and solvent environments are fast relative to the metal-molecule electron exchange processes. When $T_L=T_R=T_s$ Eqs (\ref{5}), (\ref{6}) are reduced to the standard Marcus-Hush-Chidsey expressions for electron-electrodes transfer rates \cite{27,62}. In further analysis we assume that the molecule is symmetrically coupled to the electrodes ($\Gamma_L=\Gamma_R=\Gamma$) and, unless stated otherwise, we take $\Gamma$ as a constant independent on energy.

 Given these rates, the probabilities that the dot is in the states $a$ or $b$ at time $t$, $ P_a$ and $ P_b$, are determined by the kinetic equations:
\begin{align}
\frac{d P_a}{dt} = P_b k_{b\to a} - P_a k_{a\to b};   \qquad
\frac{d P_b}{dt} = P_a k_{a \to b} - P_b k_{b\to a}       \label{7}
\end{align}  
where $ k_{a \to b} = k_{a\to b}^L + k_{a\to b}^R; \
k_{b \to a} = k_{b\to a}^L + k_{b\to a}^R. $
The steady state probabilities $P_a^0$ and $ P_b^0$ and the steady state electron current $I_{ss}$ (positive when electrons go from left to right) are given by:
\be
P_a^0 = \frac{k_{b\to a}}{k_{a\to b} + k_{b\to a}};  \qquad
P_b^0 = \frac{k_{a\to b}}{k_{a\to b} + k_{b\to a}};   \label{8}
\ee  

\begin{align}
I_{ss} = & k^L_{b\to a}P_b^0-k^L_{a\to b}P^0_a=-(k^R_{b\to a}P_b^0-k^R_{a\to b}P^0_a )
\nn\\
= & \frac{k_{a\to b}^R k_{b \to a}^L - k_{b \to a}^R k_{a \to b}^L}{(k_{a \to b} + k_{b \to a}) }.  \label{9} 
\end{align}      
\begin{figure}[t] 
\begin{center}
\includegraphics[width=8cm,height=6cm]{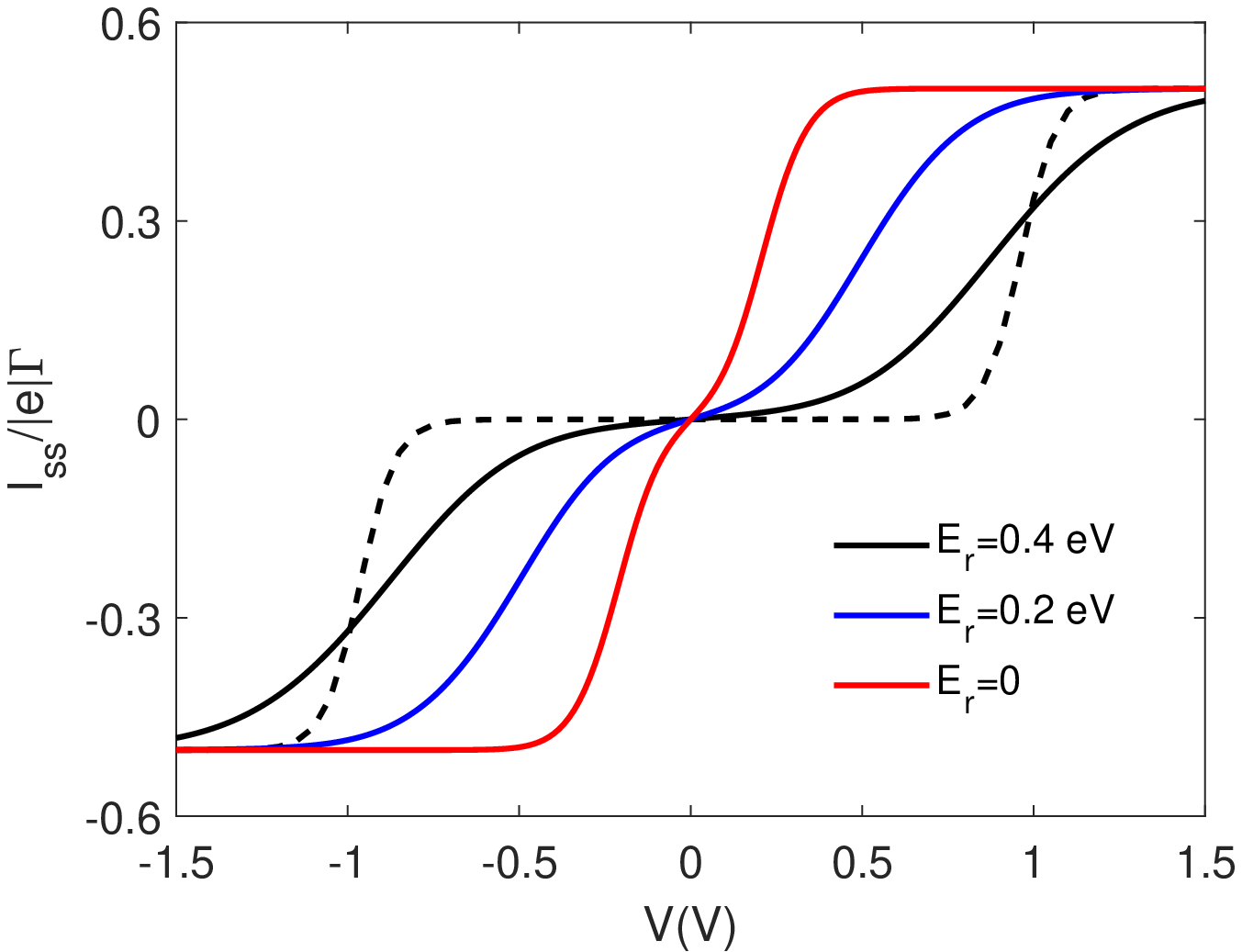} 
\includegraphics[width=8cm,height=6cm]{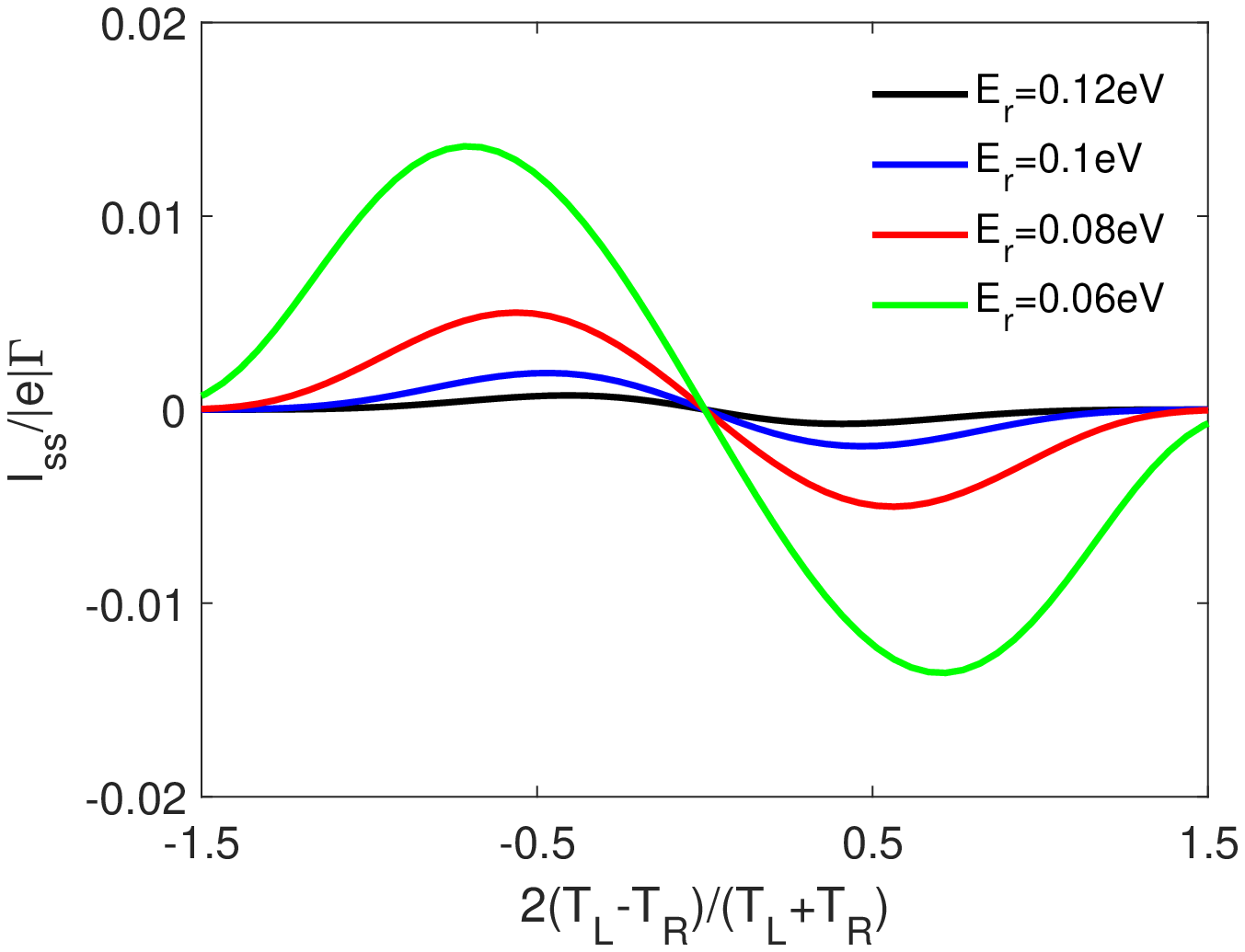}  
\caption{ Left panel: Current-voltage characteristics computed using Eqs.(\ref{10})-(\ref{12}) (solid lines). The electrodes Fermi energy in the unbiased junction is set to 0, the bias is applied symmetrically ($\mu_{L,R}=\pm|e|V/2$) relative to this origin and $T_L=T_R=T_s$. The Landauer-Buttiker limit is represented by the red line ($E_r=0$). The difference between the results obtained using Eqs.(\ref{10})-(\ref{12}) and the Marcus limit is demonstrated by comparison of the solid black line with the dashed line plotted using the Marcus equations for the electron transfer rates at the same value of $E_r$ ($E_r=0.4 eV$). Right panel: Electron current as a function of the temperature difference symmetrically distributed between the electrodes ($\ds (T_L+T_R)/2=T_s$) in an unbiased SMJ in the Marcus limit. Curves are plotted assuming that $ kT_s = 0.026 eV, \hbar\Gamma=0.01 eV$, $\epsilon_{d}=0.1 eV$ (left panel) and $\epsilon_{d}=-0.02 eV$ (right panel).
}
\label{rateI}
\end{center}\end{figure}
 
The coupling $\Gamma$of the molecular bridge to electrodes affects the electron transfer rates (and, consequently the SMJ transport properties) in two ways. First, as indicated above, it controls the transfer rates between the electrodes and the molecule. This effect is accounted for within the standard Marcus theory. Secondly, it is manifested in the lifetime broadening of molecular levels, an effect disregarded by this theory. It has been suggested in the recent work \cite{38} that the Marcus expressions for the transfer rates  may be generalized to include the broadening effect. For a symmetrically coupled system the transfer rates may be approximated as follows \cite{38,39}:
\be
k_{a \to b}^{ L,R}= \frac{ \Gamma}{\pi} \int_{-\infty}^\infty d \epsilon [1 - f_{L,R} (\beta_{L,R}, \epsilon)] K_{-}(\epsilon)  \label{10},
\ee
\be
k_{b \to a}^{L,R} = \frac{ \Gamma}{\pi}
\int_{-\infty}^\infty d \epsilon f_{L,R} (\beta_{L,R}, \epsilon)K_{+}(\epsilon),   \label{11}
\ee
where
\be
K_{\pm}(\epsilon)=Re\bigg[\sqrt\frac{\pi\beta_s}{4E_r}\exp\left[- \frac{\beta_s}{4E_r} (\hbar\Gamma\mp i(\epsilon_{d} \pm E_r - \epsilon))^2 \right]\times\mbox {erfc}(\sqrt\frac{\beta_s}{4E_r} (\hbar\Gamma\mp i(\epsilon_{d} \pm E_r - \epsilon)))\bigg]    \label{12}
\ee
and $\mbox{erfc(x)}$ is the complimentary  error function. Although the derivation of Eqs.(\ref{10})-(\ref{12}) involves some fairy strong assumptions \cite{38,39}, the result is attractive for its ability to yield the Landauer cotunneling expression in the strong molecule-electrodes coupling limit $\sqrt{E_rkT_s}\ll\hbar\Gamma$ and the Marcus expression in the opposite limit. This is shown in the left panel of Fig.1 where three current-voltage curves are plotted at the same value of $\Gamma$ and several values of $E_r$. The Landauer-Buttiker behavior is demonstrated for $E_r=0$. As $E_r$ enhances, the current-voltage curves behavior becomes more similar to the Marcus behavior represented by the dashed line. In the case of Marcus limit one observes a well pronounced plateau in the $I-V$ profile around $V=0$, as seen in the Fig.1 (dashed line). This plateau develops gradually as the electron-phonon coupling increases, and is a manifestation of a Franck-Condon blockade similar to that resulting from interactions between electrons and individual molecular vibrational modes \cite{19,20}.

When the two electrodes in an unbiased SMJ ($\mu_L=\mu_R=\mu)$ are kept at different temperatures, a thermally induced charge current emerges, as shown in Fig.1 (right panel). The current does not appear if $\epsilon_d=\mu=0$ for in this case the electron current is completely counterbalanced by the hole current. However, when $\epsilon_{d} \neq \mu $ the current emerges. The current changes its direction at  $ T_L = T_R $. Its magnitude strongly depends on the reorganization energy. Indeed, the thermally induced current takes on noticeable values only provided that the effects of nuclear reorganization are weak, and becomes suppressed when the interaction with the solvent environment increases.     

\subsection{B. Heat currents and energy conservation}

The results summarized above were mostly obtained before in works that investigate the implication of Marcus kinetics for the steady state conduction properties of molecular junctions in the limit of hopping conduction. Here we focus on the energy balance associated with such processes, and the implication of Marcus kinetics on heat transfer. Each electron hopping event between the molecule and an electrode is accompanied by solvent and metal relaxation, therefore by heat production in these  environments. We denote these heats $Q_s$ and $Q_e$ for the solvent and the electrode, respectively. Specifically, $Q_{s,a\to b}^{L,R}$ denotes the heat change in the solvent when an electron hops from the molecule into the left (L) or right (R) electrode, and similarly $Q_{s,b\to a}^{L,R}$ is heat change in the solvent in the opposite process of electron moving from the electrode to the molecule. For symmetrically coupled electrodes ($\Gamma_L=\Gamma_R=\Gamma$) considered in the Marcus limit these terms have the form :

\begin{align}
Q_{s,a\to b}^{K} = & \frac{\Gamma}{k_{a\to b}^{K}} \sqrt{\frac{\beta_s}{4 \pi E_r}} \int d \epsilon \big[1 - f_{K} (\beta_{K},\epsilon) \big] (\epsilon_{d}-\epsilon )  
\nn\\ & \times
\exp \left[- \frac{\beta_s}{4E_R} (E_{r} - \epsilon_{d}+ \epsilon)^2 \right].   \label{13}
\end{align}
and
\begin{align}
Q_{s,b \to a}^{K} = & \frac{\Gamma}{k_{b\to a}^{K}} \sqrt{\frac{\beta_s}{4 \pi E_r}} \int d \epsilon f_{K} (\beta_{K},\epsilon) ( \epsilon - \epsilon_{d}) 
\nn \\ & \times
\exp \left[- \frac{\beta_s}{4E_R} ( \epsilon_{d} + E_r - \epsilon)^2 \right].   \label{14}
\end{align}
where $K=\{L,R\}$. 
Similarly, $Q_{e,a\to b}^{K}$ and $Q_{e,b\to a}^{K}$ are heats generated in electrode K when an electron leaves (enters) the molecule into (from) that c electrode: 
\begin{align}
Q_{e,a\to b}^{K} = & \frac{\Gamma}{k_{a\to b}^{K}} \sqrt{\frac{\beta_s}{4 \pi E_r}} \int d \epsilon \big[1 - f_{K} (\beta_{K},\epsilon) \big] (\epsilon -\mu_{K})  
\nn\\ & \times
\exp \left[- \frac{\beta_s}{4 E_r} (E_{r} - \epsilon_{d} + \epsilon)^2 \right].   \label{15}
\end{align}
and
\begin{align}
Q_{e,b \to a}^{K} = & \frac{\Gamma}{k_{b\to a}^{K}} \sqrt{\frac{\beta_s}{4 \pi E_r}} \int d \epsilon f_{K} (\beta_{K},\epsilon) (\mu_{K}- \epsilon ) 
\nn\\ & \times
\exp \left[- \frac{\beta_s}{4E_R} ( \epsilon_{d} + E_r - \epsilon)^2 \right].   \label{16}
\end{align}
Eqs.(\ref{15}) and (\ref{16}) are analogs of the corresponding results reported in Ref.\cite{37}.
 
 Eqs.(\ref{13})-(\ref{16}) are expressions for the heat changes per specific hopping events. The corresponding heat change rates (heat per unit time) in the solvent and the electrodes are obtained from:
\be
J_s\equiv\dot{Q}_s=P^0_{a}(k_{a\to b}^{L}Q_{s,a\to b}^{L}+k_{a\to b}^{R}Q_{s,a\to b}^{R})+P^0_{b}(k_{b\to a}^{L}Q_{s,b\to a}^{L}+k_{b\to a}^{R}Q_{s,b\to a}^{R})        \label{17}
\ee   

 and:
\be
J^{K}_{e}\equiv\dot{Q}_{e}^{K}=k_{a\to b}^{K}P^0_{a}Q_{e,a\to b}^{K} +k_{b\to a}^{K}P^0_{b}Q_{e,b\to a}^{K}.
\label{18}
\ee
Using Eqs.(\ref{5}), (\ref{6}) and Eqs.(\ref{13})-(\ref{16}), it can be easily established that Eqs.(\ref{17}), (\ref{18}) imply:
\be
J^{L}_{e}+J^{R}_{e}+J_{s}=(\mu_L-\mu_R)I_{ss}.     \label{19}
\ee
showing the balance between heat change rates in the solvent and the electrodes and the heat generated by the current flow across the voltage bias. In the absence of solvent reorganization $J_s=0$ and Eq.(\ref{19}) is reduced to the standard junction energy balance relation $\ds J^{L}_e+\ds J^{R}_e=\ds(\mu_L-\mu_R)I_{ss}/|e|$. From Eqs.(\ref{13})-(\ref{18}) we obtain after some algebra (see Appendix A):
\be
J_{e}^{L}=(\mu_{L}-\epsilon_{d})I_{ss}-(P_{a}^{0}k_{a\to b}^{L}+P_{b}^{0}k_{b\to a}^{L})E_{r}-Y_{L} \label{20}
\ee
\be
J_{e}^{R}=(\epsilon_{d}-\mu_{R})I_{ss}-(P_{a}^{0}k_{a\to b}^{R}+P_{b}^{0}k_{b\to a}^{R})E_{r}-Y_{R} \label{21}
\ee
\be
J_{s}=2\frac{k_{a\to b}k_{b\to a}}{k_{a\to b}+k_{b\to a}}E_{r}+Y_{L}+Y_{R}   \label{22}
\ee   
where:
\begin{align}
Y_{K} =& \Gamma\sqrt{\frac{E_{r}}{\pi\beta_s}} \int d \epsilon \frac{\partial f_{K}}{\partial\epsilon}  
\nn\\ & \times
\bigg(P_{b}^{0}\exp \left[- \frac{\beta_s}{4E_R} ( \epsilon_{d} + E_r - \epsilon)^2 \right]+P_{a}^{0}\exp \left[- \frac{\beta_s}{4E_R} (E_{r}- \epsilon_{d} +\epsilon)^2 \right]\bigg)    \label{23}
\end{align}   

\begin{figure}[t] 
\begin{center}
\includegraphics[width=8cm,height=6cm]{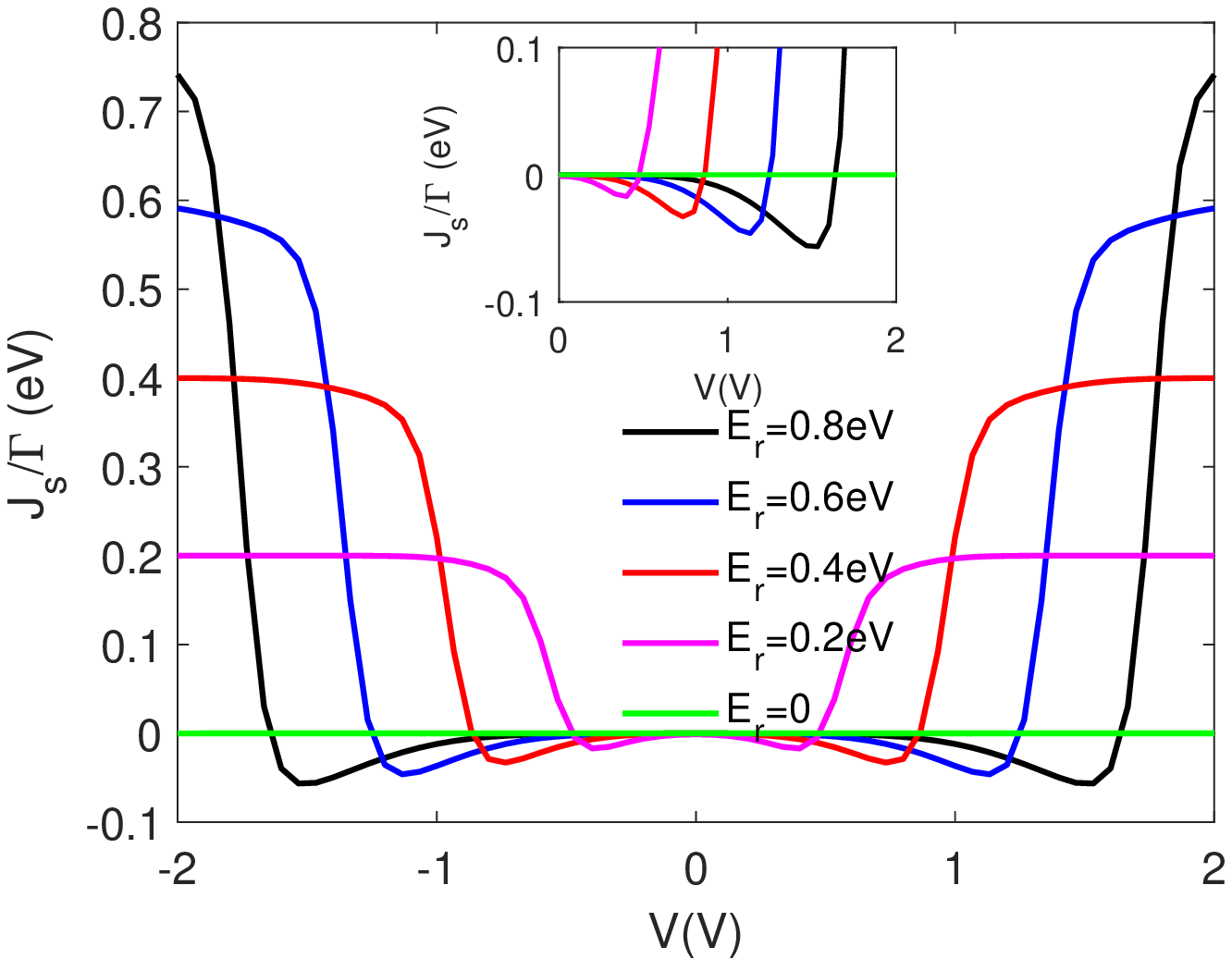} 
\includegraphics[width=8cm,height=6cm]{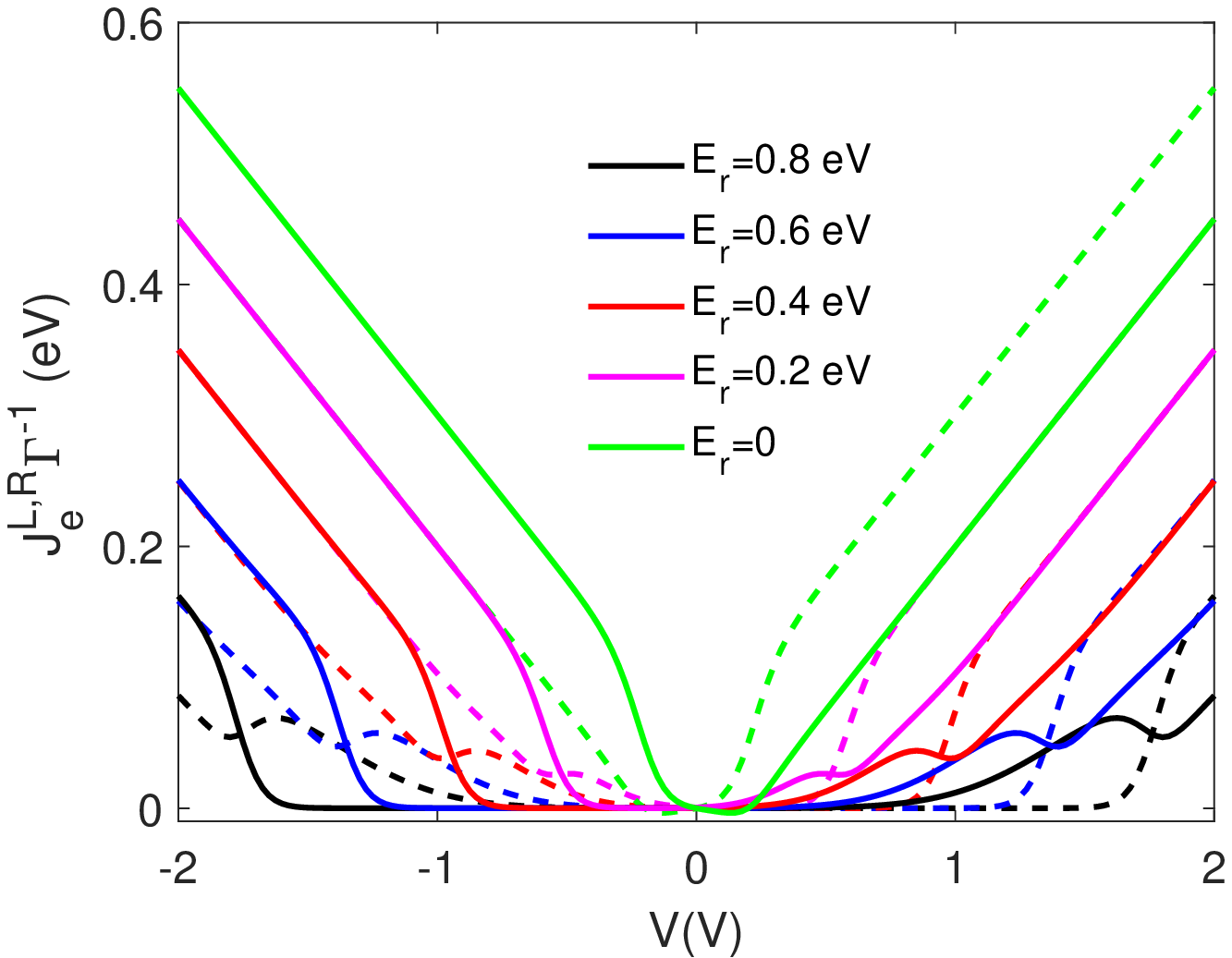}
\caption{ Heat currents $J_{s}$ (left panel), $J_{e}^{L}$ (solid lines, right panel ) and $J_{e}^{R}$ (dashed lines, right panel) shown as functions of the bias voltage $V$ for several values of the reorganization energy. Curves are plotted omitting  the effect of molecular level broadening and assuming $kT_s=kT_L=kT_R=0.026eV$, $\hbar\Gamma=0.01 eV$, $\epsilon_{d}=0.1 eV$. The inset in the left panel focuses on a segment of the main plot that emphasizes the local cooling of the solvent in the corresponding voltage range.
}
 \label{rateI}
\end{center}\end{figure}

 The analysis presented in this section remains valid regardless of the specific form of the expressions for the relevant heat flows.These may be defined within Marcus theory by Eqs.(\ref{13})-(\ref{16}) or by the generalized expressions derived using the approximation of Ref.\cite{38}:
\be
Q_{s,a\to b}^{L,R} =\frac{1}{\pi}  \frac{\Gamma}{k_{a\to b}^{L,R}} \int d \epsilon \big[1 - f_{L,R} (\beta_{L,R},\epsilon) \big] (\epsilon_d-\epsilon )K_{-}(\epsilon)      \label{24}
\ee

\be
Q_{s,b \to a}^{L,R} = \frac{1}{\pi} \frac{\Gamma}{k_{b\to a}^{L,R}}  \int d \epsilon f_{L,R} (\beta_{L,R},\epsilon) ( \epsilon - \epsilon_d) K_{+}(\epsilon)
   \label{25}
\ee
\be
Q_{e,a\to b}^{L,R} = \frac{1}{\pi} \frac{\Gamma}{k_{a\to b}^{L,R}}\int d \epsilon \big[1 - f_{L,R} (\beta_{L,R},\epsilon) \big] (\epsilon -\mu_{L,R}) K_{-}(\epsilon)      \label{26} 
\ee
 \be
Q_{e,b \to a}^{L,R} = \frac{1}{\pi}  \frac{\Gamma}{k_{b\to a}^{L,R}}  \int d \epsilon f_{L,R} (\beta_{L,R},\epsilon) (\mu_{L,R}- \epsilon ) K_{+}(\epsilon)        \label{27}
\ee
where  $k_{a\to b}^{L,R}$, $k_{b\to a}^{L,R}$ and $K_{\pm}(\epsilon)$ are given by Eqs(\ref{10})-(\ref{12}). It should be noted that the procedure leading to Eq.(\ref{19}) that demonstrates the energy conservation remains the same when these expressions are used.

Results based on Eqs.(\ref{20})-(\ref{24}) are displayed in Figure 2. The left panel shows the heat deposited in the solvent environment plotted against the bias voltage for different values of the reorganization energy. The right panel shows similar results for the left and right electrodes. The following observations can be made:

(a). Reflecting the behavior of the electronic current, energy exchange processes are very weak at low bias due to the Franck-Condon blockade that hinders electron transport. Noticeable heat currents appear when $|e|V$ exceeds the reorganization energy $E_r$ thus lifting the blockade.

(b). The heat deposited into the electrodes shows an asymmetry between positive and negative biases (or equivalently between left and right electrodes). This asymmetry reflects the different positioning of the energy of the transferred electron relative to the left and right Fermi energies \cite{65}, and was observed experimentally \cite{66}.

(c). The heat deposited into the solvent environment (left panel) is symmetric with respect to bias inversion because it reflects energy balance relative to both electrodes.

(d). Note that the heat exchanged with the solvent (nuclear) environment can become negative, namely, heat may be pulled out of this environment at some range of bias and reorganization energy. In the present case of a symmetrically coupled SMJ with a symmetrically distributed bias voltage this happens at $|eV|\approx 2E_r$, namely when the driving force originating from the bias is nearly counterbalanced by forces originating from elecron-phonon interactions. This cooling is reminiscent of similar effects discussed in the low electron-phonon coupling regime \cite{59,60,61}.

\subsection{III. Driven junction}

Next we consider charge and energy currents in driven biased junctions, where driving is modeled by an externally controlled time dependent parameter in the system Hamiltonian. In the present study we limit our consideration to time dependence of the single electron "level" $\epsilon_d$ of the molecular bridge that may in principle be achieved by varying the gate potential. Similar studies in the absence of electron-phonon interactions, focusing on a consistent quantum thermodynamic description of such systems were recently published \cite{67,68,69,70,71,72,73,74,75,76,77,78,79,80,81}. The model considered below includes strong coupling to the phonon environment at the cost of treating this coupling semiclassically and assuming weak coupling between molecule and electrodes. This model is similar to that used to analyze cyclic voltammetry observations when extended to consider two metal interfaces \cite{82}.

In further analysis we assume that the electron level $ \epsilon_d$ is a slowly varying and construct an expansion of the solution of in powers of $\dot\epsilon_d$ \cite{83}. To this end we start by separating the time dependent populations into their steady state components (which implicitly depend on time through $\epsilon_d(t)$ and corrections defined by \cite{29}:
\be
P_a(t) = P_a^0 (\epsilon_d) - G (\epsilon_d, t);  \qquad
P_b(t) = P_b^0 (\epsilon_d) + G(\epsilon_d, t).  \label{28} 
\ee
where the steady state populations $P^0_{a,b}$ satisfy $P_bk_{b\to a}-P_ak_{a\to b}=0$; $P_a+P_b=1$ and are given by Eq.(\ref{8}). Then:
\be
\frac{dP_a}{dt} = \dot{\epsilon_d} \frac{\partial P_a^0}{\partial \epsilon_d} - \frac{d G}{d t};  \qquad
\frac{d P_b}{dt} = \dot{\epsilon_d} \frac{\partial P_b^0}{\partial \epsilon_d} + \frac{d G}{d t}.  \label{29}
\ee
From Eqs.(\ref{28}) it follows that
\be
\frac{dP_a}{dt} = -\frac{dP_b}{dt}=G(k_{a\to b}+k_{b\to a}).  \label{30}
\ee         
Comparing (\ref{29}) and (\ref{30}) we obtain:
\be
\frac{dG}{dt}  = \dot{\epsilon_d} \frac{\partial P_a^0}{\partial \epsilon_d} - G (k_{a \to b} + k_{b \to a}).  \label{31}  
\ee    
The electronic currents can be written in terms of $G$ in the form:
\begin{align} 
 I_L=k^L_{b\to a} P_b-k^L_{a\to b} P_a=I_{ss}+I_L^{excess};    \qquad  I_L^{excess}=(k^L_{b\to a}+k^L_{a\to b})G
\nn\\
I_R=k^R_{b\to a} P_b-k^R_{a\to b} P_a=I_{ss}+I_R^{excess};   \qquad   I_R^{excess}=(k^R_{b\to a}+k^R_{a\to b})G         \label{32}
\end{align}
expressing the fact that the left and right excess particle (electron) current due to driving are generally not the same. Eqs.(\ref{32}) together with analogs of (\ref{17}) and (\ref{18}) in which $P^0_{a,b}$ are replaced by $P_{a,b}$ can be used to obtain the excess heat currents caused by driving
\be
J^{excess}_{s}=\dot{Q}^{excess}_{s}=G(k^L_{b\to a}Q^L_{s,b\to a}-k^L_{a\to b}Q^L_{s,a\to b}+k^R_{b\to a}Q^R_{s,b\to a}-k^R_{a\to b}Q^R_{s,a\to b})  \label{33}
\ee
\be
J^{K,excess}_{e}=\dot{Q}^{K,excess}_{e}=G(k^K_{b\to a}Q^K_{e,b\to a}-k^K_{a\to b}Q^K_{e,a\to b})         \label{34}
\ee
Using these expressions with Eqs.(\ref{13})-(\ref{16}) for the heat currents and Eqs.(\ref{5}),(\ref{6}) (or (\ref{10}), (\ref{11})) we find:
\be
 J^{excess}_{tot}\equiv J^{excess}_{s}+J^{L,excess}_{e}+J^{R,excess}_{e}=G[(\mu_L-\epsilon_d)(k^L_{a\to b}+k^L_{b\to a})+(\mu_R-\epsilon_d)(k^R_{a\to b}+k^R_{b\to a})]     \label{35}
\ee
Eqs.(\ref{31})-(\ref{35}) are exact relations. In particular, Eq.(\ref{31}) can be used as a basis for expansions in powers of $\dot\epsilon_d$. We start by writing $G$ as such a power series: $G=G^{(1)}+G^{(2)}+...$ with $G^{(n)}$ representing order $n$ in $\dot\epsilon_d$ and use this expansion in Eq.(\ref{31}) while further assuming that $G$ depends on time only through its dependence on $\epsilon_d$ implying that $\ds dG_{n}/dt$ is of order $n+1$. We note that our results are consistent with this assumption.

\subsection{A The quasistatic limit: First order corrections}.

To first order in$\dot\epsilon_d$ the left hand side of Eq.(\ref{31}) vanishes, leading to 
\be
G^{(1)} = \dot{\epsilon_d}\frac{\partial P_a^0}{\partial \epsilon_d} \frac{1}{(k_{a \to b} + k_{b \to a})}.  \label{36}
\ee     
where $k_{a\to b}$ and $k_{b\to a}$ depend on time through their dependence on $\epsilon_d$. Note that Eqs.(\ref{30}) and (\ref{36}) imply that $dP^{(1)}_a/dt=-dP^{(1)}_b/dt=\dot\epsilon_d\partial P^0_a/\partial\epsilon_d$, namely this order of the calculation corresponds to the quasistatic limit where all dynamics is derived from the time dependence of $\epsilon_d$. At the same time it should be emphasized that this limit is not a reflection of the instantaneous steady state, as is evident from Eqs.(\ref{32}).

Consider first the electronic current. Using Eqs.(\ref{32}) and (\ref{36}) the first order correction to the electron exchange rates with the left and right electrodes is obtained in the form:
\be
I^{(1)}_K=(k^K_{b\to a}+k^K_{a\to b})G^{(1)}=\dot\epsilon_d\frac{\partial P^0_a}{\partial\epsilon_d}\nu_K     \label{37}
\ee
with
\be
\nu_K=\frac{k^K_{a\to b}+k^K_{b\to a}}{k_{a\to b}+k_{b\to a}}              \label{38}
\ee
namely, a product of the (first order) change in the electronic population on the molecule $dP^{(1)}_a/dt$ and the fraction $\nu_K$ of this change associated with the electrode $K$.

Next consider the heat currents. Using Eqs.(\ref{36}) for $G$ in Eq.(\ref{35}) leads to:
\be
J_{tot}^{(1)}=-\epsilon_d\dot\epsilon_d\frac{\partial P_a^0}{\partial\epsilon_d}+\dot\epsilon_d\frac{\partial P_a^0}{\partial\epsilon_d}\Bigg(\mu_L\frac{k_{a\to b}^{L}+k_{b\to a}^{L}}{k_{a\to b}+k_{b\to a}}+ \mu_R\frac{k_{a\to b}^{R}+k_{b\to a}^{R}}{k_{a\to b}+k_{b\to a}}\Bigg).  \label{39}
\ee 
To better elucidate the physical meaning of this result we rearrange the first term on the right according to $-\ds\epsilon_d\dot\epsilon_d\partial P^0_a/\partial\epsilon_d=\dot\epsilon_d P^0_a-d(\epsilon_d P^0_a)/dt$ and use Eq.(\ref{37}) to cast Eq.(\ref{39}) in the form:
\be
   \frac{d(\epsilon_d P^0_a)}{dt}\equiv \dot{E}^{(1)}_M =\dot\epsilon_d P^0_a-J^{(1)}_{tot}+\mu_L\dot{n}_L+\mu_R\dot{n}_R      \label{40}
\ee    
 This equation is a statement of the first law of thermodynamics, where $\dot{E}^{(1)}_M$ represents to order $1$, the rate of change of energy in the molecule and the terms on the right stand for the work per unit time ($\dot\epsilon_d P^0_a$), rate of heat developing in the environment ($-J^{(1)}_{tot}\equiv -(J^{L(1)}_e+J^{R(1)}_e+J^{(1)}_s))$ and rate of chemical work ($\mu_L I^{(1)}_L+\mu_R I^{(1)}_R$) to the same order.. All terms included in Eq.(\ref{40}) are of the form $\dot\epsilon_dr(\epsilon_d)$ where $r$ is an arbitrary function and are therefore the same except of sign when $\epsilon_d$ goes up or down, as this should be within the quasistatic regime. 

\subsection{B. Beyond the quasistatic regime: Second order corrections}

Using the expansion for $G$ in Eq.(\ref{31}) and keeping only second order terms leads to: 
\be
G^{(2)}=-\frac{1}{k_{a\to b}+k_{b\to a}}\frac{dG^{(1)}}{dt}           \label{41}
\ee
The second order correction to the electron current is (K=\{L,R\}):
which, using Eq.(\ref{36}) gives:
\be
G^{(2)}=-\frac{\dot\epsilon_d^2}{(k_{a\to b}+k_{b\to a})}\frac{\partial}{\partial\epsilon_d} \left[\frac{\partial P_a^0}{\partial\epsilon_d}\frac{1}{(k_{a \to b} + k_{b \to a}}\right]       \label{42}
\ee
The second order correction to the electron current is (K=\{L,R\}):
\be
I^{(2)}_K=(k^K_{a\to b}+k^K_{b\to a})G^{(2)}=-\dot\epsilon_d^2\frac{\partial}{\partial\epsilon_d}\left[\frac{\partial P^0_a}{\partial\epsilon_d}\frac{1}{(k_{a\to b}+k_{b\to a})}\right]\nu_K.                            \label{43}
\ee 
The second order excess heat is obtained from Eq.(\ref{35}) by replacing $G$ with $G^{(2)}$. The sum of second order corrections to the heat currents then takes the form: 
\begin{figure}[t] 
\begin{center}
\includegraphics[width=8.5cm,height=6cm]{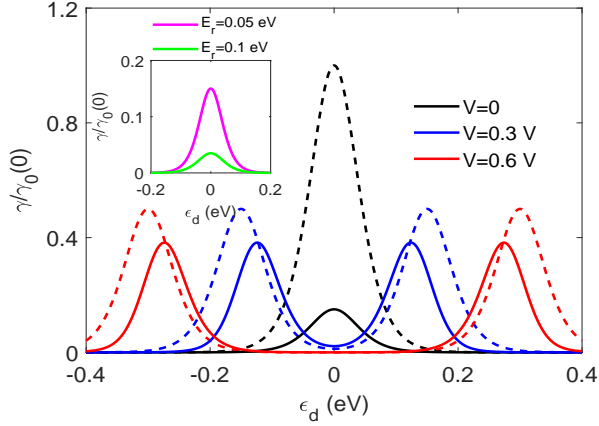} 
\caption{ Friction coefficient characterizing dissipation in the system caused by driving  of the energy level in a symmetrically coupled junction with a symmetrically applied bias as a function of $\epsilon_d$. Main figure: Dashed lines represent friction in the absence of molecule-solvent coupling ($E_r$=0); full lines are plotted at $E_r=0.05 eV$ for three values of the bias voltage (indicated by different colors). Inset: Friction at zero bias for the indicated two values of reorganization energy. In all lines, the friction is normalized by its value at zero bias and zero reorganization energy. Other parameters are: $kT_L= kT_R=kT_s = 0.026 eV,$, $\hbar\Gamma =0.01 eV$.
}
 \label{rateI}
\end{center}\end{figure}
The sum of second order corrections to the heat currents then takes the form:
\be
J^{(2)}_{tot}\equiv J_{e}^{L(2)}+J_{e}^{R(2)}+J_{s}^{(2)}=-\epsilon_d G^{(2)}(k_{a\to b}+k_{b\to a})+G^{(2)}\Big(\mu_L(k_{a\to b}^{L}+k_{b\to a}^{L})+ \mu_R(k_{a\to b}^{R}+k_{b\to a}^{R})\Big).  \label{44}
\ee 
Using Eqs.(\ref{42}), (\ref{44}) we can present the work-energy balance equation at this order as follows:
\be
\dot{E}^{(2)}_M=\dot{W}^{(2)}-J^{(2)}_{tot}+(\mu_L I^{(2)}_L+\mu_R I^{(2)}_R)    \label{45}
\ee
where
\be
\dot E^{(2)}_M=-\frac{d}{dt}\big[\epsilon_d G^{(1)}\big]=-\dot\epsilon_d^2\Bigg[\frac{\partial P^0_a}{\partial\epsilon_d}\frac{1}{k_{a\to b}+k_{b\to a}}+\epsilon_d\frac{d}{d\epsilon_d}\Bigg(\frac{\partial P^0_a}{\partial\epsilon_d}\frac{1}{k_{a\to b}+k_{b\to a}}\Bigg)\Bigg]           \label{46}
\ee
is the second order change rate in the total system energy expressed as the time derivative of the first order contribution to this energy (product of $\epsilon_d$ and the first order correction to the population $G^{(1)}$), and
\be
\dot{W}^{(2)}=-\dot\epsilon_d G^{(1)}\equiv-\dot\epsilon_d^{2}\frac{\partial P_{a}^0}{\partial\epsilon_d}\frac{1}{k_{a\to b}+k_{b\to a}}  \label{47}
\ee
is the second order excess work per unit time (power) which corresponds to the lowest order {\it irreversible} work expressing dissipation caused by driving the level. The last term on the right hand side of (\ref{45}) represents the second order contribution to the rate of chemical work, thus Eq.(\ref{45}) is an expression for the first law of thermodynamics at the second order of our expansion. Following Refs\cite{67,84}, the coefficient in front of $\dot\epsilon_d^2$ in Eq.(\ref{47})
\be
	\gamma=-\frac{\partial P^0_a}{\partial\epsilon_d}\frac{1}{k_{a\to b}+k_{b\to a}}        \label{48}
	\ee
may be identified with the friction coefficient. Similar interpretation was suggested in Refs.\cite{67,84} for different models for SMJs.

 Dependencies of $\gamma$ on $\epsilon_d$ are shown in Fig.3. In an unbiased junction the friction coefficient reaches its maximum at $\epsilon_d=\mu=0$ and falls down approaching zero as $\epsilon_d$ moves away from this position. In this case $\gamma$ appears to increase with the increasing voltage. This results from the fact that at low bias the Franck-Condon blockade discussed above makes molecule-electrodes coupling small, and friction increases  upon removing this blockade at higher bias voltage. Also, at higher voltage the peak splits -two peaks appear due to electron-electrodes  exchange near the two Fermi energies characterizing the biased junction. Note that coupling to the solvent shifts the positions of these peaks, in correspondence with the Eqs.(\ref{5}), (\ref{6}) for transfer rates.  
\subsection{C. Evolution of the system (dot) entropy}

Define the system entropy by the Gibbs formula for our binary system
\be
S=-k(P_a\ln P_a+(1-P_a)\ln (1-P_a))        \label{49}
\ee
Using Eqs.(\ref{28}) we find
\be
S=-k((P_a^0-G)\ln( P_a^0-G)+(1-P_a^0+G)\ln (1-P_a^0+G))         \label{50}
\ee
which can be used to find again an expansion in powers of $\dot\epsilon_d$: $S=S^{(0)}+S^{(1)}+...$. In what follows we limit ourselves to the case of an unbiased junction in the wide band limit for which $P^0_a/P^0_b=\ds\exp(-\beta\epsilon_d)$. In the absence of driving
\be
S^{(0)}=-k(P_a^{(0)}\ln P_a^{(0)}+(1-P_a^{(0)})\ln (1-P_a^{(0)}))     \label{51}
\ee
and (assuming that $T_L=T_R=T_s\equiv T$)
\be
S^{(1)}=-k\beta\epsilon_dG^{(1)}.                     \label{52}
\ee
The first and second order variations in the dot's entropy due to driving are obtained as (recall that the sign of $J_{tot}$ was chosen so that the heat current into  the environment is positive):
\be
\dot{S}^{(1)}=\dot\epsilon_d\frac{\partial S^{(0)}}{\partial\epsilon_d}=\frac{1}{T}\epsilon_d\dot\epsilon_d\frac{\partial P^{(0)}_a}{\partial\epsilon_d}=-\frac{J^{(1)}_{tot}}{T}                                             \label{53}
\ee
and
\be
\dot{S}^{(2)}=\dot\epsilon_d\frac{\partial S^{(1)}}{\partial\epsilon_d}=-k\beta\dot\epsilon_dG^{(1)}-k\beta\epsilon_d\dot\epsilon_d\frac{\partial G^{(1)}}{\partial\epsilon_d}=\frac{\dot{W}^{(2)}}{T}-\frac{J^{(2)}}{T}.           \label{54}
\ee
Eq.(\ref{54}) may be rewritten as:
\be
\dot{S}^{(2)}+\frac{J^{(2)}}{T}=\frac{\dot{W}^{(2)}}{T}.            \label{55}
\ee
The left side of Eq.(\ref{55}) is the sum of the rate of total entropy change in the system (dot/molecule) $\dot{S}^{(2)}$ and the entropy flux into the electrodes and solvent environment. Together these terms give the total entropy production due to the irreversible nature of the process at this order. This result is identical to that obtained in fully quantum mechanical treatments of similar processes evaluated in the absence of coupling to solvent \cite{67,78,80}, except of a sign difference in the heat definition. Here, the heat which is going {\it out} of the system is defined as positive.

\subsection{IV. Marcus junction engine}

In this Section, we extend the above analysis to discuss a simple model that simulates an atomic scale engine. This can be achieved by imposing asymmetry on the coupling of the molecular bridge to the electrodes that enables to convert the motion of $\epsilon_d$ to electron current between the electrodes. A simple choice is:
\be
\Gamma_{L,R}(\epsilon)=\Gamma\frac{\delta^2}{(\epsilon \pm \epsilon_0)^2+\delta^2}.       \label{56}
\ee
where, for definiteness, we assign the ($+$) sign to the left electrode. This represents a situation where the moving level is coupled to wide-band electrodes via single level gateway sites with energies $\pm\epsilon_0$ attached to the left/right electrode. The electron transfer rates are calculated from Eqs.(\ref{5}), (\ref{6}). In further calculations we assume that $\epsilon_d$ varies according to
\be
\epsilon_d(t)= E_0-E_1\cos(2\pi t/\tau)                \label{57}
\ee
It is intuitively obvious that fast enough driving (small $\tau$) with a choice of origin $E_0$ and amplitude $E_1$ that encompass the interval ($-\epsilon_0,\epsilon_0$) will produce current from the left to the right electrode which may be appreciable if $\epsilon_0$ is sufficiently larger than $T_L=T_R=T_s\equiv T$. This current is given by the average over a period:
\be
{<I>}_{\tau}=\frac{1}{\tau}\int_{0}^{\tau}dtI_L(t)=\frac{1}{\tau}\int_{0}^{\tau}dtI_R(t).            \label{58}
\ee
where $I_K(t)$ are given by Eq.(\ref{32}). Further analytical progress can be made by using the expansion in powers of $\dot\epsilon_d$. However, using this expansion implies that $\delta$ in Eq.(\ref{56}) is large enough for the inequality $kT\Gamma(\epsilon)\gg\dot\epsilon_d$ to be satisfied for all $\epsilon$. The lowest non-vanishing contribution to Eq.(\ref{58}) is then:
\be
{<I>}^{(2)}_{\tau}=\frac{1}{\tau}\int_{0}^{\tau}dtI^{(2)}_L(t)=\frac{1}{\tau}\int_{0}^{\tau}dtI^{(2)}_R(t).    \label{59}
\ee 
where the second order contributions to the currents are given by Eq.(\ref{43}). Note that this is the excess current produced by driving which persists also in the absence of imposed bias.
\begin{figure}[t] 
\begin{center}
\includegraphics[width=8.5cm,height=6cm]{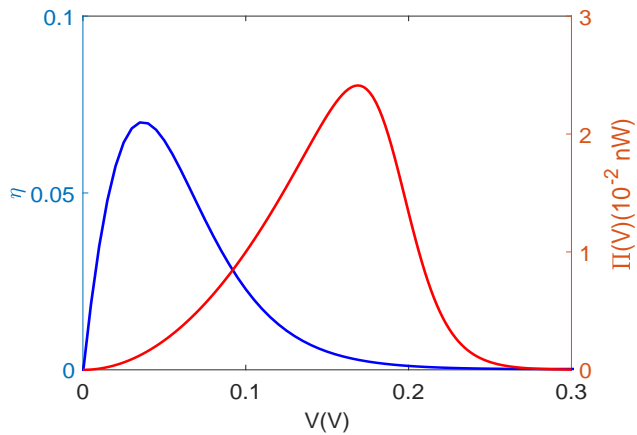} 
\caption{  The averaged over the period thermodynamic efficiency $\eta$ (blue line) and power $\Pi$ (red line)produced in the junction by periodically driving the bridge level. Curves are plotted at $kT=0.026 eV$, $\hbar\Gamma=0.01 eV$, $E_0=0$, $E_1=0.2 eV$, $E_r=0.05 eV$, $\tau=10 ps$.
}
\label{rateI}
\end{center}\end{figure}

When a voltage bias is imposed so as to drive a current in the opposite direction to ${<I>}^{(2)}_{\tau}$, the total current
\be
{<I>}_{\tau}(V)=I_{ss}(V)+{<I>}^{(2)}_{\tau}(V)          \label{60}
\ee
can be used to define the power produced by the engine:
\be
\Pi(V)=V(I_{ss}(V)+{<I>}^{(2)}_{\tau}(V)).               \label{61}
\ee
The device efficiency is defined as the ratio:
\be
\eta(V)=\frac{\Pi(V)}{\ds\frac{1}{\tau}\int_{0}^{\tau}dt \dot{W}^{(2)}(V,t)}        \label{62}
\ee
Figure 4. shows the voltage dependence of these engine characteristics. Obviously both vanish in the absence of load ($V=0$) as well as at the stopping voltage when the current vanishes, and go through their maxima at different 'optimal' voltages (which in turns depend on the choices of $E_0$ and $E_1$). Note that because of the intrinsic friction in this model, the efficiency vanishes rather than maximizes at the stopping voltage point.



\subsection {V. Conclusions}

In the present work we have analyzed energy balance in single-molecule junctions characterized by strong electron-phonon interactions, modeled by a single level molecule (dot) connecting free electron metal electrodes, where charge transfer kinetics is described by Marcus electron transfer theory. The standard steady state transport theory was extended to include also slow driving of the molecular level that may be achieved by employing a time dependent gate potential. A consistent description of the energetics of this process was developed leading to the following observations:

(a) Accounting for the total energy and its heat, work and chemical components shows that energy conservation (first law of thermodynamics) is satisfied by this model at all examined order of driving.

(b) Heat is obviously produced by moving charge across potential bias. In addition, when charge transfer involves solvent reorganization, the current flowing in a biased junction can bring about heat transfer between the metal and the solvent environments, and may even produce solvent cooling in some voltage range.

(c) In the presence of solvent reorganization the friction experienced by the driven coordinate $\epsilon_d$ which expresses energy loss (heat production) due to the molecule-metal electron exchange is strongly affected by the presence of solvent reorganization.

(d) Beyond the reversible (driving at vanishingly small rate) limit, entropy is produced and is determined, at least to the second order in the driving speed, by the excess work associated with the friction affected by the molecule coupling to the electrodes and solvent environments.

We have also used this model to study a molecular junction with a periodically modulated dot energy. We have considered a model engine in which such periodic driving with a properly chosen energy depended molecule-electrode coupling can move charge against a voltage bias and calculated the power and efficiency of such a device. In the parameter range consistent with our mathematical modeling useful work can be produced only in the irreversible regime, and we could determine the points of optimal performance of such engine with respect to power and efficiency.

While our calculations are based on Marcus electron transfer kinetics in which level broadening due to molecule-metal coupling is disregarded, we have shown that extension to the more general kinetics suggested in Ref.\cite{38}, which (approximately) bridges between Marcus sequential hopping and Landauer cotunneling limit is possible.

Energy conversion on the nanoscale continues to be focus of intense interest. The present calculation provides a first simple step in evaluating such phenomena in a system involving electron transport, electron-solvent interaction and mechanical driving.

\subsection{Acknowledgments}

The present work was supported by the U.S National Science Foundation (DMR-PREM 1523463). Also, the research of AN is supported by the Israel-U.S Binational Science Foundation, the German Research Foundation (DFG TH 820/11-1), the U.S National Science foundation (Grant No.CHE1665291) and the University of Pennsylvania. NZ acknowledges support of the Sackler Visiting Professor Chair at Tel Aviv University, Israel.

\subsection{Appendix A}

Here, we derive Eq.(\ref{20}) for $\dot{Q}^L_e$. Starting from Eq.(\ref{15}) we present $Q^L_{e,a\to b}$ in the form:
\be
Q^L_{e,a\to b}=\epsilon_d-E_r-\mu_L+\frac{\Gamma}{k_{a\to b}^{L}} \sqrt{\frac{\beta_s}{4 \pi E_r}} \int d \epsilon \big[1 - f_{L} (\beta_{L},\epsilon) \big] (\epsilon +E_r-\epsilon_d)\exp \left[- \frac{\beta_s}{4 E_r} (E_{r} - \epsilon_{d} + \epsilon)^2 \right].   \label{63}  
\ee
Similarly:
\be
Q^L_{e,b\to a}=\mu_L-\epsilon_d-E_r+\frac{\Gamma}{k_{b\to a}^{L}} \sqrt{\frac{\beta_s}{4 \pi E_r}} \int d \epsilon  f_{L} (\beta_{L},\epsilon) (\epsilon_d +E_r-\epsilon)\exp \left[- \frac{\beta_s}{4 E_r} (E_{r} +\epsilon_{d} - \epsilon)^2 \right].   \label{64}  
\ee
Integrating by parts we obtain:
\be
Q^L_{e,a\to b}=\epsilon_d-E_r-\mu_L-\frac{\Gamma}{k_{a\to b}^{L}}\sqrt{\frac{E_r}{\pi\beta_s}}\int d\epsilon\frac{\partial f_L(\beta_L,\epsilon)}{\partial\epsilon}
\exp \left[- \frac{\beta_s}{4 E_r} (E_{r} -\epsilon_{d} +\epsilon)^2 \right].     \label{65}
\ee
and
\be
Q^L_{e,b\to a}=\mu_L-\epsilon_d-E_r-\frac{\Gamma}{k_{b\to a}^L}\sqrt{\frac{E_r}{\pi\beta_s}}\int d\epsilon\frac{\partial f_L(\beta_L,\epsilon)}{\partial\epsilon}
\exp \left[- \frac{\beta_s}{4 E_r} (E_{r} +\epsilon_{d} - \epsilon)^2 \right].     \label{66}
\ee
Substituting these expressions into Eq.(\ref{18}) we get the expression for $\dot{Q}^L_e$ given by Eq.(\ref{20}). Expressions for $\dot{Q}^R_e$ and $\dot{Q}^R_e$ and $\dot{Q}_s$ may be derived in the same way


\begin{thebibliography}{99}
\bibitem{83} The smallness of $\dot\epsilon_d$ should be established by comparison to $kT\Gamma$ and $\Gamma^2/\hbar$ \cite{67}.
\end{thebibliography}
\end{document}